\documentclass[11pt]{article}

\usepackage{geometry}
\usepackage[utf8]{inputenc}
\usepackage[T1]{fontenc}
\usepackage{amsmath,amssymb,amsfonts}	
\usepackage{graphicx}
\usepackage{color}
\usepackage{xcolor}
\usepackage{xspace}
\usepackage{booktabs}
\usepackage{slashed}
\usepackage{tablefootnote}
\usepackage{makecell}
\usepackage[normalem]{ulem}

\usepackage[colorlinks=true,pdfstartview=FitV, linkcolor=blue, citecolor=blue,urlcolor=blue]{hyperref} 
\usepackage[sort&compress,numbers,colon,merge]{natbib}

\makeatletter
\g@addto@macro\bfseries{\boldmath}
\makeatother

\usepackage[retainorgcmds]{IEEEtrantools}

\newenvironment{eqns*}
{\begin{IEEEeqnarray*}{+rCl"s+x*}}
{\end{IEEEeqnarray*}}

\begin{document}
\begin{titlepage}
\vspace*{-1cm}
\phantom{hep-ph/***}
\flushright
\hfil{CPPC-2023-06}

\vskip 1.5cm
\begin{center}
\mathversion{bold}
{\LARGE\bf
    When Energy Goes Missing: \\[1ex]
    New Physics in $b\to s \nu\nu$ with Sterile Neutrinos
}\\[3mm]
\mathversion{normal}
\vskip .3cm
\end{center}
\vskip 0.5  cm
\begin{center}
{\large Tobias Felkl}$^{1}$,
{\large Anjan Giri}$^{2}$,
{\large Rukmani Mohanta}$^{3}$,
{\large Michael A.~Schmidt}$^{1}$
\\
\vskip .7cm
{\footnotesize
$^{1}$ Sydney Consortium for Particle Physics and Cosmology, School of Physics, The University of New South Wales, Sydney, NSW 2052, Australia\\[0.3cm]
$^{2}$ Department of Physics, IIT Hyderabad, Kandi - 502285, India\\[0.3cm]
$^{3}$ School of Physics, University of Hyderabad, Hyderabad - 500046, India\\[0.3cm]

\vskip .5cm
\begin{minipage}[l]{.9\textwidth}
\begin{center}
\textit{E-mail:}
\tt{t.felkl@unsw.edu.au},
\tt{giria@phy.iith.ac.in},
\tt{rmsp@uohyd.ac.in},
\tt{m.schmidt@unsw.edu.au}
\end{center}
\end{minipage}
}
\end{center}
\vskip 1cm
\begin{abstract}
Belle II recently reported the first measurement of $B^+\to K^+ + \mathrm{inv}$, which is $2.8\sigma$ above the Standard Model prediction. We explore the available parameter space of new physics within Standard Model effective field theory extended by sterile neutrinos ($\nu$SMEFT) and provide predictions for the other $B\to K^{(\star)} +\mathrm{inv}$ decay modes and invisible $B_s$ decays. We also briefly comment on charged current decays $B\to D^{(\star)} \ell \bar \nu$ and possible ultraviolet completions of the relevant $\nu$SMEFT operators.
\end{abstract}
\end{titlepage}


\section{Introduction}
\label{sec:intro}

The Belle II experiment recently announced the first measurement of $B^+\to K^+ +\mathrm{inv}$ with a branching ratio of  $\mathrm{BR}(B^+\to K^+ +\mathrm{inv}) = (2.4\pm 0.7)\times 10^{-5}$~\cite{Glazov:2023EPS}, which is $2.8\sigma$ above the Standard Model (SM) prediction BR($B^+\to K^+ \nu\bar\nu)_{\rm SM}=(5.06\pm0.14\pm0.28)\times 10^{-6}$~\cite{Becirevic:2023aov}. 
The result assumes 3-body decay kinematics as predicted by the SM and is based on two independent analyses, one inclusive analysis and one with a hadronic tag. While the hadronic tag analysis is consistent with the SM prediction within $1\sigma$, the more sensitive inclusive tag analysis shows a $3\sigma$ deviation from the SM prediction. Together with the previous searches by Belle~\cite{Belle:2013tnz,Belle:2017oht}, BaBar~\cite{BaBar:2013npw} and Belle II~\cite{Belle-II:2021rof}, a simple weighted average results in BR$(B^+\to K^++\mathrm{inv}) = (1.4\pm0.4)\times 10^{-5}$~\cite{Glazov:2023EPS}. 
Motivated by the result, Ref.~\cite{Bause:2023mfe,Allwicher:2023xyz} studied the implication within Standard Model effective field theory (SMEFT) and Ref.~\cite{Athron:2023hmz} demonstrated that non-universal lepton-flavour-conserving $Z'$ models are not able to accommodate the excess and thus proposed the introduction of lepton-flavour-violating couplings. Several earlier works~\cite{Browder:2021hbl,He:2021yoz,Felkl:2021uxi,He:2022ljo,Ovchynnikov:2023von,Asadi:2023ucx} also proposed explanations for an excess in $B^+\to K^++\mathrm{inv}$ motivated by the result of the 2021 Belle II analysis~\cite{Belle-II:2021rof}, which first used the inclusive tagging method.

One attractive explanation of the excess in $B^+\to K^++\mathrm{inv}$ is an additional decay channel with undetected final states, like sterile neutrinos~\cite{He:2021yoz,Felkl:2021uxi}, dark matter~\cite{Bird:2004ts,Altmannshofer:2009ma,He:2022ljo} or more generally long-lived particles~\cite{Berezhiani:1989fs,Berezhiani:1990jj,Berezhiani:1990wn,Filimonova:2019tuy,Ferber:2022rsf,Ovchynnikov:2023von}. 
Additional light sterile neutrinos are particularly well motivated, and occur in numerous minimal extensions of the SM, see e.g.~\cite{Iguro:2018qzf,Azatov:2018kzb,Asadi:2018sym,Babu:2018vrl,Heeck:2018ntp,Gomez:2019xfw}. Light sterile neutrinos have been proposed as an explanation of the deviation from lepton flavour universality in $R(D^{(\star)})$~\cite{Ligeti:2016npd,Asadi:2018sym,Robinson:2018gza,Bardhan:2019ljo,Balaji:2019kwe,Mandal:2020htr,Du:2022ipt,Datta:2022czw} by introducing new decay channels, $B\to D^{(\star)}\tau N$, with sterile neutrinos $N$.
Light GeV-scale sterile neutrinos may also explain neutrino masses via the seesaw mechanism~\cite{Minkowski:1977sc} like in the $\nu$MSM~\cite{Asaka:2005pn}. The lifetime of sterile neutrinos is constrained from big bang nucleosynthesis (BBN) to be shorter than $0.02$ s~\cite{Boyarsky:2020dzc}. 
Together with searches for sterile neutrinos at beam dump experiments, this translates into a lower bound on their mass in the minimal seesaw model of $\mathcal{O}(350)$ MeV apart from a small mass window at 120--140 MeV~\cite{Bondarenko:2021cpc}. Although the lower bound is well below the $B^+$ meson mass, the sterile neutrino mass will become important for a relevant part of the parameter space and thus should be included in the analysis.

$B$ meson decays are best described within low energy effective theory (LEFT). Most studies focus on vector operators~\cite{Colangelo:1996ay,Melikhov:1998ug,Altmannshofer:2009ma,Buras:2014fpa}, because they are predicted in the SM, but sterile neutrinos motivate to extend the analysis to scalar and tensor operators. The scenario of massless sterile neutrinos has been discussed in~\cite{Kim:1999waa,Aliev:2001in,Browder:2021hbl,Bause:2021cna,Bause:2023mfe} and the general case with massive neutrinos has been considered by some of us in~\cite{Felkl:2021uxi} in terms of helicity amplitudes~\cite{Gratrex:2015hna}. There are also recent studies of $B\to K^{(\star)} \nu\bar \nu$ decays within SMEFT, which discuss the complementarity with other observables~\cite{Bause:2020auq,Bause:2021cna,Bause:2023mfe} including comments on scalar operators with (massless) right-handed neutrinos~\cite{Bause:2021cna,Bause:2023mfe}.

In this work, we will build on a previous analysis within LEFT~\cite{Felkl:2021uxi}, but work within SM effective field theory extended by sterile neutrinos ($\nu$SMEFT). Within the framework of $\nu$SMEFT we consider additional decay channels to sterile neutrinos including the full sterile neutrino mass dependence and identify the viable regions of parameter space which explain the excess in $B^+\to K^++\mathrm{inv}$ and are consistent with the non-observation of the other three $B\to K^{(\star)} +\mathrm{inv}$ decay modes, invisible $B_s$ decays and measurements of charged current $B\to D^{(\star)}\ell \bar\nu$ decays. We also point out ultraviolet (UV) completions for the viable $\nu$SMEFT operators.

The remainder of the paper is structured as follows. In Sec.~\ref{sec:EFT} we introduce the effective field theory (EFT) framework, discuss both $\nu$SMEFT and LEFT and how they are connected via renormalisation group (RG) running. The relevant observables are introduced in Sec.~\ref{sec:pheno} and the results discussed in Sec.~\ref{sec:results}. Finally, possible UV completions are presented in Sec.~\ref{sec:uv_completions} before concluding in Sec.~\ref{sec:conclusions}. A few technical details are relegated to the Appendix.


\section{Effective field theory framework}
\label{sec:EFT}

Starting at dimension-6, there are operators in $\nu$SMEFT which contribute to $b\to s\nu\nu$ at tree level. We focus on the four semi-leptonic 4-fermion operators with sterile neutrinos~\cite{Liao:2016qyd}
\begin{equation}\label{eq:Lagrangian}
\begin{aligned}
  \mathcal{L}  & = C^{\rm QN} (\bar Q \gamma_\mu Q) (\bar N \gamma^\mu N)
    + C^{\rm dN} (\bar d \gamma_\mu d) (\bar N\gamma^\mu N)
    \\
    &
    + C^{\rm LNQd} (\bar L^\alpha N)\epsilon_{\alpha\beta} (\bar Q^\beta d)
    + C^{\rm LNQdT} (\bar L^\alpha\sigma^{\mu\nu}N) \epsilon_{\alpha\beta} (\bar Q^\beta\sigma_{\mu\nu} d)\;,
\end{aligned}%
\end{equation}
where $Q$, $L$ and $d$ are the SM left-handed quark doublet, left-handed lepton doublet and right-handed down-type quark singlet, respectively, and $N$ denotes right-handed neutrinos, i.e.~right-handed SM singlet fermions.
Flavour and colour indices are suppressed, $\sigma^{\mu\nu}=\tfrac{i}{2}[\gamma^\mu,\gamma^\nu]$ and the Levi-Civita tensor is defined with $\epsilon_{12}=1$. Assigning lepton number $+1$ to $N$, all four operators are lepton number conserving.
In contrast to~\cite{Liao:2016qyd}, we introduce a tensor operator. This operator basis is convenient as $\mathcal{O}^{\rm LNQd}$ and $\mathcal{O}^{\rm LNQdT}$ do not mix under 1-loop QCD renormalisation group (RG) running. The tensor operator $\mathcal{O}^{\rm LNQdT}$ is related to the scalar operators in~\cite{Liao:2016qyd} via a Fierz transformation
\begin{equation}
     (\bar L^\alpha \sigma^{\mu\nu} N)\epsilon_{\alpha\beta}(\bar Q^\beta \sigma_{\mu\nu} d)
     = 
     -8(\bar L^\alpha d)\epsilon_{\alpha\beta} (\bar Q^\beta N)   -4 (\bar L^\alpha N) \epsilon_{\alpha\beta} (\bar Q^\beta d)\;.
\end{equation}
For this analysis we define all $\nu$SMEFT operators at the scale $\mu=1$ TeV. At the electroweak scale $\Lambda_{\rm EW} = m_Z$, the $\nu$SMEFT operators are matched onto LEFT.

Only a few operators in LEFT are generated at tree level from the set of $\nu$SMEFT operators introduced above. The relevant interactions for $b\to s \nu \nu$ 
and $b\to c \ell \nu$ processes are described by the Lagrangian~\cite{Aebischer:2017gaw,Jenkins:2017jig}
\begin{equation}
\begin{aligned}
    \mathcal{L} 
    =
    \sum_{X=L,R}  
    C^{\text{VLX}}_{\nu d} \mathcal{O}^{\text{VLX}}_{\nu d}
    + 
    \Big(&
    C_{\nu d}^{\text{SLL}} \mathcal{O}_{\nu d}^{\text{SLL}} + C_{\nu d}^{\text{TLL}} \mathcal{O}_{\nu d}^{\text{TLL}} 
  \\&+
    C_{\nu edu}^{\text{VLL}} \mathcal{O}_{\nu edu}^{\text{VLL}} + 
    C_{\nu edu}^{\text{SLL}} \mathcal{O}_{\nu edu}^{\text{SLL}} + C_{\nu edu}^{\text{TLL}} \mathcal{O}_{\nu edu}^{\text{TLL}}+\mathrm{h.c.}
    \Big)
    \end{aligned}
\end{equation}
with the effective operators
\begin{equation}
\begin{aligned}
    \mathcal{O}_{\nu d}^{\text{VLX}}  & = ( \overline{\nu_L}\gamma_\mu  \nu_L)(\overline{d_X} \gamma^\mu  d_X)\;, 
                                      &
        \mathcal{O}_{\nu d}^{\text{SLL}} 
    & =  ( \overline{\nu^c_L}  \nu_L)(\overline{d_R}  d_L)\;,
    &
    \mathcal{O}_{\nu d}^{\text{TLL}} & 
    = (\overline{\nu^c_L} \sigma_{\mu\nu}  \nu_L)  (\overline{d_R} \sigma^{\mu\nu}  d_L)\;, 
    \\
    \mathcal{O}_{\nu e du}^{\text{VLL}}  & = ( \overline{\nu_L}\gamma_\mu  e_L)(\overline{d_L} \gamma^\mu  u_L)\;,
                                         &
    \mathcal{O}_{\nu edu}^{\text{SLL}} 
    & =  ( \overline{\nu^c_L}  e_L)(\overline{d_R}  u_L)\;,
    &
    \mathcal{O}_{\nu e du}^{\text{TLL}} & 
    = (\overline{\nu^c_L} \sigma_{\mu\nu}  e_L)  (\overline{d_R} \sigma^{\mu\nu}  u_L)
    \;.
\end{aligned}
\end{equation}
Right-handed neutrinos $N$ are expressed in terms of left-handed Weyl spinors $\nu_L = N^c \equiv C \overline{N} ^T$ with the charge conjugation matrix $C=i\gamma^2\gamma^0$. SM neutrinos carry the generation index $\alpha=1,2,3$ and sterile neutrinos are labelled with $\alpha\geq 4$.

Note that the scalar operator $\mathcal{O}_{\nu d}^{\text{SLL}}$ is symmetric in the neutrino flavours and the tensor operator $\mathcal{O}_{\nu d}^{\text{TLL}}$ is antisymmetric in the neutrino flavours, which can be straightforwardly derived from 
\begin{equation}
\begin{aligned}\label{eq:symmetry-properties}
    \overline{\psi_i^c} \Gamma\psi^{cj} &= \eta_\Gamma \overline{\psi^j} \Gamma \psi_i\;,
    \;\;
    C^{-1}\Gamma C = \eta_\Gamma \Gamma^T \;,
                                        &
    \eta_\Gamma & = \begin{cases} +1 & \mathrm{for}\;\;\Gamma=1,\gamma_5,\gamma^\mu\gamma_5\\
    -1 & \mathrm{for} \;\; \Gamma = \gamma^\mu ,\sigma^{\mu\nu}, \sigma^{\mu\nu}\gamma_5
    \end{cases}
    \;.
\end{aligned}
\end{equation}
The other operators do not exhibit any manifest symmetry properties.\footnote{There are further LEFT operators
\begin{equation}
\begin{aligned}
    \mathcal{O}_{\nu d}^{\text{SLR}} & = (\overline{\nu^c_L}\nu_L)(\overline{d_L}d_R) 
&
    \mathcal{O}_{\nu e du}^{\text{VLR}} & = (\overline{\nu_L}\gamma_\mu e_L)(\overline{d_R}\gamma^\mu u_R) 
    \\
     \mathcal{O}_{\nu edu}^{\text{SRL}} & = (\overline{\nu_L}e_R)(\overline{d_R}u_L) 
    & \mathcal{O}_{\nu e du}^{\text{VRL}} & = (\overline{\nu^c_L}\gamma_\mu e_R)(\overline{d_L}\gamma^\mu u_L) 
   \\
    \mathcal{O}_{\nu edu}^{\text{SLR}} & = (\overline{\nu^c_L}e_L)(\overline{d_L}u_R) 
   &
   \mathcal{O}_{\nu e du}^{\text{VRR}} & = (\overline{\nu^c_L}\gamma_\mu e_R)(\overline{d_R}\gamma^\mu u_R)
    \\
    \mathcal{O}_{\nu edu}^{\text{SRR}} & = (\overline{\nu_L}e_R)(\overline{d_L}u_R)
&
    \mathcal{O}_{\nu e du}^{\text{TRR}} & = (\overline{\nu_L}\sigma_{\mu\nu}  e_R)(\overline{d_L}\sigma^{\mu\nu}u_R)
\end{aligned}
\end{equation}
that also contribute to bespoke processes, but are not induced via tree-level matching to the considered dimension-6 $\nu$SMEFT operators. They are however referenced in Appendix~\ref{sec:SPVAT} where the mapping onto the $S,P,V,A,\mathcal{T}$ basis is presented.
}

The matching conditions have been obtained by translating the existing matching results in the literature~\cite{Jenkins:2017jig,Li:2019fhz,Li:2020lba} to the operator basis we are using. 
All LEFT operators are given in the mass eigenstate basis, while the operators in $\nu$SMEFT are defined in the interaction basis, where the charged leptons and down-type quarks are mass eigenstates. Up-type quark mass eigenstates $\hat u_i$ and neutrino mass eigenstates $\hat \nu_i$ are related to the interaction basis by
\begin{align}
  u_\alpha & = (V^\dagger)_{\alpha i} \hat u_i \;, &
  \nu_\alpha & = U_{\alpha i} \hat \nu_i
  \;,
\end{align}
where $V$ denotes the Cabibbo-Kobayashi-Maskawa (CKM) matrix, and $U$ is the leptonic mixing matrix. The top-left $3\times 3$ sub-block of the leptonic mixing matrix $U$ is approximately given by the Pontecorvo-Maki-Nakagawa-Sakata (PMNS) matrix. For a large part of the parameter space, (sterile) neutrino masses are negligible compared to the $B$ meson mass, $m_\alpha\ll m_B$, and thus (sterile) neutrinos can be treated as massless and the PMNS matrix is unphysical in this limiting case. Even for sterile neutrinos with GeV-scale masses, CHARM~\cite{CHARM:1985nku}, NuTeV~\cite{NuTeV:1999kej} and DELPHI~\cite{DELPHI:1996qcc} constrain the active-sterile mixing to be small, $|U|^2\lesssim 10^{-5}$, for the relevant region in parameter space. Thus, in the following we neglect leptonic mixing and treat the sterile neutrinos as mass eigenstates, and so drop the hats from the mass eigenstates for simplicity.

At tree level, we find the following non-vanishing down-quark flavour-violating Wilson coefficients for $b\to s\nu\nu$ processes and vector operators with $\alpha,\beta\geq 4$
\begin{equation}
    \begin{aligned}
    C_{\nu d,\alpha \beta sb}^{\text{VLL}} &= 
        -C^{\rm Q N}_{sb (\beta-3)(\alpha-3) } 
    \;,
                                           &
    C_{\nu d,\alpha\beta sb}^{\text{VLR}} &= 
        -C^{\rm d N}_{sb (\beta-3)(\alpha-3)} \;, 
\end{aligned}
\end{equation}
and for the scalar and tensor operators
\begin{equation}\label{eq:STop}
    \begin{aligned}
    C^{\text{SLL}}_{\nu d, \alpha \beta sb} &= 
    \begin{cases} 
    \frac12 C^{\rm L N Qd*}_{\beta(\alpha-3) bs}
    & \alpha\geq4;\;
    1\leq\beta\leq 3
    \\
    \frac12 C^{\rm L N Qd*}_{\alpha(\beta-3) bs}
    & 1\leq\alpha\leq 3;\; \beta\geq 4
    \end{cases}\;,
                                            &                                     
         C^{\text{SLL}}_{\nu d, \alpha \beta bs} &= 
    \begin{cases} 
    \frac12 C^{\rm L N Qd*}_{\beta(\alpha-3) sb}
    & \alpha\geq4;\;
    1\leq\beta\leq 3
    \\
    \frac12 C^{\rm L N Qd*}_{\alpha(\beta-3) sb}
    & 1\leq\alpha\leq 3;\; \beta\geq 4
    \end{cases}\;,
                                             \\
    C^{\text{TLL}}_{\nu d, \alpha \beta sb} &=
    \begin{cases} 
\frac12 C^{\rm LNQdT*}_{\beta(\alpha-3) bs}
    & \alpha\geq4;\; 1\leq\beta\leq 3
    \\
- \frac12 C^{\rm LNQdT*}_{\alpha(\beta-3) bs}
    & 1\leq\alpha\leq 3;\; \beta\geq 4
    \end{cases}\;.
                                            &
   C^{\text{TLL}}_{\nu d, \alpha \beta bs} &=
    \begin{cases} 
\frac12 C^{\rm LNQdT*}_{\beta(\alpha-3) sb}
    & \alpha\geq4;\; 1\leq\beta\leq 3
    \\
- \frac12 C^{\rm LNQdT*}_{\alpha(\beta-3) sb}
    & 1\leq\alpha\leq 3;\; \beta\geq 4
    \end{cases}\;.
\end{aligned}
\end{equation}
Note, we explicitly (anti-)symmetrised the scalar (tensor) Wilson coefficients.
Furthermore, at tree level the non-vanishing LEFT operators 
for $b\to c\ell\nu$ processes are given by 
scalar and tensor operators which are matched to $\nu$SMEFT operators
\begin{equation} 
\begin{aligned}
    C^{\text{SLL}}_{\nu edu, \alpha \beta bc} &=
        -V_{cm}^*C^{\rm L N Qd*}_{\beta(\alpha-3) mb} \;, 
                                              & 
    C^{\text{TLL}}_{\nu edu, \alpha \beta bc} &=
        - V_{cm}^*C^{\rm LNQdT*}_{\beta(\alpha-3) mb}
        & \mathrm{with}\, &\,\, \alpha\geq 4
        \;.
  \end{aligned}
  \end{equation}
In addition to the contribution from $\nu$SMEFT dimen\-sion-6 operators, the SM contribution to the LEFT operators relevant for $b\to s,c$ processes can be summarised as
\begin{equation}
\begin{aligned}
  C^{\rm VLL, SM}_{\nu d, \alpha\beta sb} & = - \frac{4 G_F}{\sqrt{2}} \frac{\alpha}{2\pi} V_{ts}^* V_{tb} \frac{X}{\sin^2\theta_w}\delta_{\alpha\beta}
  \;,
                                          &
 C^{\rm VLL,SM}_{\nu e du, \alpha\beta bc} & = -\frac{4 G_F}{\sqrt{2}} V_{cb}^*\delta_{\alpha\beta}
\end{aligned}
\end{equation}
in terms of the function $X$~\cite{Buras:1998raa}.
The Wilson coefficient for $b\to s\nu\bar\nu$ includes electroweak corrections induced by top quarks at 2-loop order, which are numerically given by $X=6.3772 \sin^2\theta_w$~\cite{Brod:2010hi,Straub:2018kue,peter_stangl_2023_7994776}.

The dominant RG corrections  originate from strong interactions. While the (axial-)vector current operators do not receive QCD corrections at 1-loop order, scalar and tensor operators receive RG corrections. The 1-loop RG equations for the scalar and tensor Wilson coefficients $C^S$ and $C^T$ are
\begin{align}
     \frac{d \ln C^S}{d\ln \mu}  &= -3C_F \frac{\alpha_s}{2\pi}\;,
     &
     \frac{d \ln C^T}{d\ln \mu}  &= C_F \frac{\alpha_s}{2\pi}\;,
\end{align}
with the second Casimir invariant $C_F=(N_c^2-1)/2N_c=4/3$ for $N_c=3$ and $\alpha_s=g_s^2/4\pi$. Taking into account 1-loop QCD running, we arrive at 
\begin{equation} 
\begin{aligned}
    C^S(\mu_1) &= \left(\frac{\alpha_s(\mu_2)}{\alpha_s(\mu_1)}\right)^{3C_F/b} C^S(\mu_2)\;,
               &
    C^T(\mu_1) &= \left(\frac{\alpha_s(\mu_2)}{\alpha_s(\mu_1)}\right)^{-C_F/b} C^T(\mu_2)\;,
\end{aligned}
\end{equation}
where $b=-11+\tfrac23 n_f$ and $n_f$ is the number of active quark flavours. As there is no operator mixing for 1-loop QCD running in the chosen operator basis, there is a one-to-one correspondence between a given LEFT Wilson coefficient at the hadronic scale $\mu=4.8$ GeV and the $\nu$SMEFT Wilson coefficient at $\mu=1$ TeV: All scalar LEFT operators are rescaled with the factor $1.608$ and tensor operators with $0.853$ compared to the result from matching $\nu$SMEFT to LEFT. 

\section{Phenomenology}
\label{sec:pheno}

Belle II measured~\cite{Glazov:2023EPS} the decay $B^+\to K^++\mathrm{inv}$ with a branching ratio of BR($B^+\to K^+ +\mathrm{inv})=(2.4\pm0.7)\times 10^{-5}$ assuming a 3-body decay with massless neutrinos, which is above the 90\% CL exclusion limit of BR($B^+\to K^+ +\mathrm{inv})<1.6\times 10^{-5}$~\cite{BaBar:2013npw}. Together with the previous searches for $B^+\to K^++\mathrm{inv}$~\cite{BaBar:2013npw,Belle:2017oht,Belle:2021idw}, the Belle II measurement results in a simple weighted average of BR($B^+\to K^++\mathrm{inv})=(1.4\pm0.4)\times 10^{-5}$~\cite{Glazov:2023EPS}. The other searches for $B\to K^{(\star)}+\mathrm{inv}$ so far only resulted in upper limits of BR($B^0\to K^0+\mathrm{inv})<2.6\times 10^{-5}$~\cite{Belle:2017oht}\footnote{Reference \cite{Belle:2017oht} quotes the upper bound on the branching ratio for $B^0\to K_S^0\nu\nu$ which we translated to $B^0\to K^0\nu\nu$.}, BR($B^0\to K^{\star 0} +\mathrm{inv})<1.8\times 10^{-5}$~\cite{Belle:2017oht}, and BR($B^+\to K^{\star +} +\mathrm{inv})<4.0\times 10^{-5}$~\cite{Belle:2013tnz}.

The short-distance contributions to the differential decay rates of semi-leptonic $B$ meson decays are straightforwardly expressed in terms of helicity amplitudes
\begin{equation} 
\begin{aligned}\label{eq:BKnunu}
    \frac{d\Gamma(\bar B\to \bar K \nu_\alpha\nu_\beta)}{dq^2} & = \frac{1}{4(1+\delta_{\alpha\beta})} G^{(0)}(q^2)\;,
                                                               &
    \frac{d\Gamma(\bar B\to \bar K^\star(\to \bar K\pi) \nu_\alpha\nu_\beta)}{dq^2} & = \frac{3}{4(1+\delta_{\alpha\beta})} G^{0,0}_0(q^2)\;,
\end{aligned}
\end{equation}
where the coefficients $G^{(0)}(q^2)$ and $G^{0,0}_0(q^2)$ of the Wigner-D functions are defined in~\cite{Gratrex:2015hna} with the replacements for neutrinos as discussed in \cite{Felkl:2021uxi} and $q^2$ denotes the missing invariant mass squared carried away by the two neutrinos.\footnote{Note, in contrast to \cite{Felkl:2021uxi}, the symmetry factor $(1+\delta_{\alpha\beta})$ for identical neutrinos has been explicitly included in the differential decay rate instead of the normalisation factor $N$.} Kinematics constrains $q^2$ to lie in the range $(m_\alpha+m_\beta)^2\leq q^2\leq (m_{B^+}-m_{K^+})^2$.
We use the recently published $B\to K^{(\star)}$ form factors from~\cite{Gubernari:2023puw} with the parametrisation defined in~\cite{Bharucha:2015bzk}.
Following~\cite{Descotes-Genon:2019bud}, we increase the $B\to K^\star\nu\nu$ branching ratios by 20\% to account for finite-width effects.

\begin{figure}[ptb!]
\centering
\includegraphics[width=0.7\linewidth]{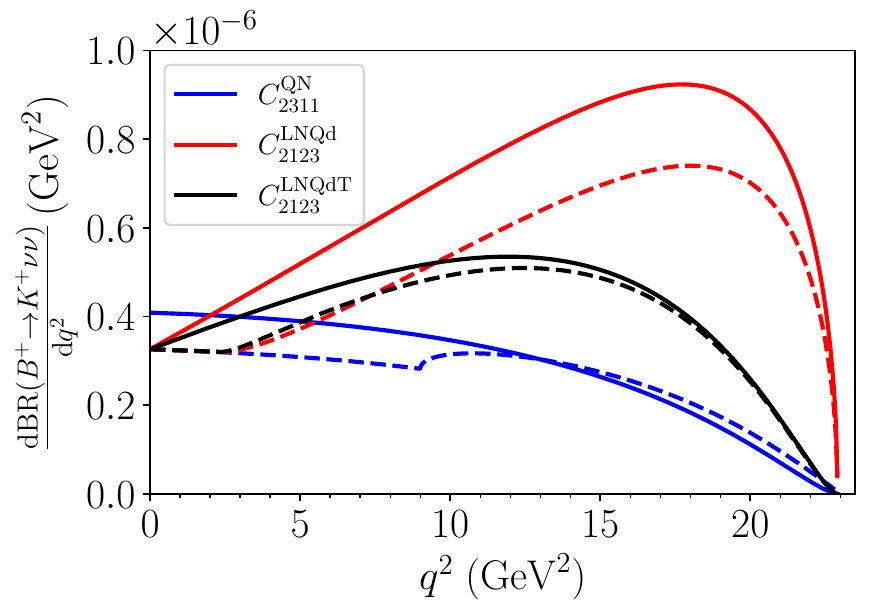}
\caption{Differential branching ratio as a function of the missing invariant mass squared $q^2$ for vector (blue), scalar (red) and tensor (black) operators. All Wilson coefficients are fixed to $C=0.01\, \mathrm{TeV}^{-2}$ at $\mu=1$ TeV and the SM contribution is taken into account. The solid lines are for massless sterile neutrinos and the dashed line for a sterile neutrino of mass $m_1=1.5$ GeV. 
}
\label{fig:diff}
\end{figure}

Note that massive sterile neutrinos and the Lorentz structure of the operators result in modifications to the $q^2$ distribution compared to the SM prediction, see e.g.~\cite{Felkl:2021uxi}, which is illustrated in Fig.~\ref{fig:diff} for $B^+\to K^+\nu\nu$. The $q^2$ distributions of $B\to K^{(\star)}\nu\nu$ decays are also dependent on the sterile neutrino mass and Lorentz structure of the operators. Heavy sterile neutrinos do not contribute below $q^2<4m_1^2$ for the vector operators $\mathcal{O}^{\rm QN}$ and $\mathcal{O}^{\rm dN}$ and $q^2<m_1^2$ for scalar operators $\mathcal{O}^{\rm LdQN}$ and tensor operators $\mathcal{O}^{\rm LdQNT}$. Below the kinematic cutoff for sterile neutrinos, the dashed lines in Fig.~\ref{fig:diff} approach the SM prediction. In comparison to vector operators, scalar and tensor operators are enhanced for large $q^2$. The experimental efficiency however is largest for small $q^2$ and reduced for large $q^2$. As there is currently no publicly available $q^2$ distribution which could be used for recasting the Belle II result~\cite{Glazov:2023EPS}, we only consider branching ratios and do not take into account modifications to the $q^2$ distribution. Viable regions of parameter space which explain the branching ratio measured by Belle II can only be taken as indicative and we would like to stress the need for a more detailed discussion of sterile neutrino final states based on the full $q^2$ distribution.

In addition to the short-distance contribution, there is a long-distance tree-level contribution to
$B^+\to K^{(\star)+} \nu\bar\nu$ in the SM.~\cite{Kamenik:2009kc} It is mediated by a $\tau$ lepton, $B^+\to  \tau^+ (\to K^{(\star)+} \bar\nu_\tau)\nu_\tau$. The interference between the long-distance and short-distance contributions is negligible due to the narrow $\tau$ resonance~\cite{Kamenik:2009kc} and thus the total branching ratio for $B^+\to K^{(\star)+} \nu\bar\nu$ is approximately obtained by summing the short- and long-distance contributions. Although the long-distance contribution is subdominant, since it only amounts to about 10\% of the short-distance contribution, we nevertheless include it in the numerical analysis.

Another relevant decay channel is the invisible decay of $B_s$ mesons. Reference \cite{Alonso-Alvarez:2023mgc} recently derived a first upper bound on the branching ratio of invisible $B_s$ decays using LEP data with $\mathrm{BR}(B_s\to \mathrm{inv}) <5.9\times 10^{-4}$. It may also
be probed at the Belle II experiment: Using an integrated luminosity of $5\,\mathrm{ab}^{-1}$ allows to probe BR($B_s\to\mathrm{inv})>1.1\times 10^{-5}$~\cite{Belle:2018ezy}. The branching ratio for identical final state neutrinos is given by 
\begin{equation}
\begin{aligned}
  \mathrm{BR}(B_s\to \nu_\alpha\nu_\alpha)  & = \frac{\tau_{B_s}m_{B_s}^3 f_{B_s}^2}{32\pi}
    \Bigg( 
    \left|C_{\nu d,\alpha\alpha sb}^{\text{VLR}}-C_{\nu d,\alpha\alpha sb}^{\text{VLL}}\right|^2 x_\alpha^2\sqrt{1-4x_\alpha^2}
    \\&
    +\left|C_{\nu d,\alpha\alpha bs}^{\text{SLL}*} -C_{\nu d,\alpha\alpha sb}^{\text{SLL}}\right|^2 \frac{(1-4x_\alpha^2)^{3/2}}{4\,x_b^2}
    +\left|C_{\nu d,\alpha\alpha bs}^{\text{SLL}*} +C_{\nu d,\alpha\alpha sb}^{\text{SLL}}\right|^2 \frac{\sqrt{1-4x_\alpha^2}}{4\,x_b^2}
    \\&
    +\mathrm{Re}\left[(C_{\nu d,\alpha\alpha sb}^{\text{VLR}}-C_{\nu d,\alpha\alpha sb}^{\text{VLL}}) 
    (C_{\nu d,\alpha\alpha bs}^{\text{SLL}} +C_{\nu d,\alpha\alpha sb}^{\text{SLL}*})\right] 
    \frac{x_\alpha \sqrt{1-4x_\alpha^2}}{x_b}
    \Bigg) \;,
\end{aligned}
\end{equation}
with $x_\alpha=m_\alpha/m_{B_s}$ and $x_b=m_b/m_{B_s}$. The general expression for different neutrino final states is presented in App.~\ref{sec:Bsnunu}. Vector contributions are helicity suppressed and thus in the limit of vanishing sterile neutrino masses, there is only a contribution from scalar operators~\cite{Bause:2021cna,Bause:2023mfe} 
\begin{equation}
\begin{aligned}
    \mathrm{BR}(B_s\to \nu_\alpha\nu_\alpha)  & = \frac{\tau_{B_s}m_{B_s}^5 f_{B_s}^2}{64\pi m_b^2} 
    \left( 
    \left|C_{\nu d,\alpha\alpha bs}^{\text{SLL}}\right|^2 + \left|C_{\nu d,\alpha\alpha sb}^{\text{SLL}}\right|^2 
    \right)\;. 
\end{aligned}
\end{equation}

Directly related to the discussed processes are the rare decays $B_s\to \phi \nu \nu$ and $\Lambda_b\to \Lambda^{(\star)} \nu\nu$ which provide independent information. The differential branching ratio for $B_s\to \phi\nu\nu$ has the same form as the one for $B\to K^*\nu\nu$ with the form factors replaced by the $B_s\to \phi$ form factors presented in~\cite{Gubernari:2023puw}. See \cite{Rajeev:2021ntt} for a calculation in SMEFT. The differential branching ratio for $\Lambda_b\to \Lambda^{(\star)} \nu\bar\nu$ has been calculated in \cite{Das:2023kch} within SMEFT. Within $\nu$SMEFT it may be obtained from the differential branching ratio for the rare $\Lambda_b$ decay with charged leptons, see~\cite{Bordone:2021usz}, similar to the discussion of $B\to K^{(\star)}\nu\nu$ in \cite{Felkl:2021uxi}. These channels are inaccessible at the Belle II experiment running at the $\Upsilon(4S)$ resonance, but a Tera-Z experiment such as FCC-ee or CEPC may be able to measure these channels as pointed out in~\cite{Li:2022tov,Amhis:2023mpj}. Given the uncertain experimental situation, we leave the discussion of these channels for future work. 
If the anomaly persists, it will be highly interesting to measure $B_s\to\phi\nu\nu$ and $\Lambda_b\to \Lambda^{(\star)} \nu\nu$ to pin down the Lorentz structure of the effective operators.

Additional constraints arise from light-lepton flavour universality ratios for charged current processes $b\to c\ell \bar\nu$. The differential branching ratios are obtained by exchanging $\nu_\alpha\to e_\alpha$ in Eq.~\eqref{eq:BKnunu} and dropping the factors $(1+\delta_{\alpha\beta})$ for identical final state particles, see also \cite{Bhattacharya:2022bdk}.
The relevant form factors for $B\to D^{(\star)}$ are taken from~\cite{Gubernari:2018wyi}. 
Three different ratios have been measured: Belle measured  $\frac{\text{BR}(B\to D\mu\nu)}{\text{BR}(B\to De\nu)}=0.995\pm0.045$~\cite{Belle:2015pkj} which is consistent with the SM prediction of $0.997$. Similarly, the measurement of $\frac{\text{BR}(B^0\to D^{\star -}e^+\nu)}{\text{BR}(B^0\to D^{\star -}\mu^+\nu)}=1.01\pm0.032$~by  Belle \cite{Belle:2018ezy} is consistent with the SM prediction of 1.005. Most recently, Belle II measured the inclusive ratio 
$\frac{\text{BR}(B\to X e\nu)}{\text{BR}(B\to X \mu \nu)}=1.007\pm0.009\pm0.019$~\cite{Belle-II:2023qyd} which also agrees with the SM prediction of $1.006\pm0.001$~\cite{Rahimi:2022vlv}. 
The deviation from unity in the three ratios originates from phase space suppression due to the muon mass. 

\begin{figure*}[tb!]
    \centering
    
    \includegraphics[width=0.5\textwidth]{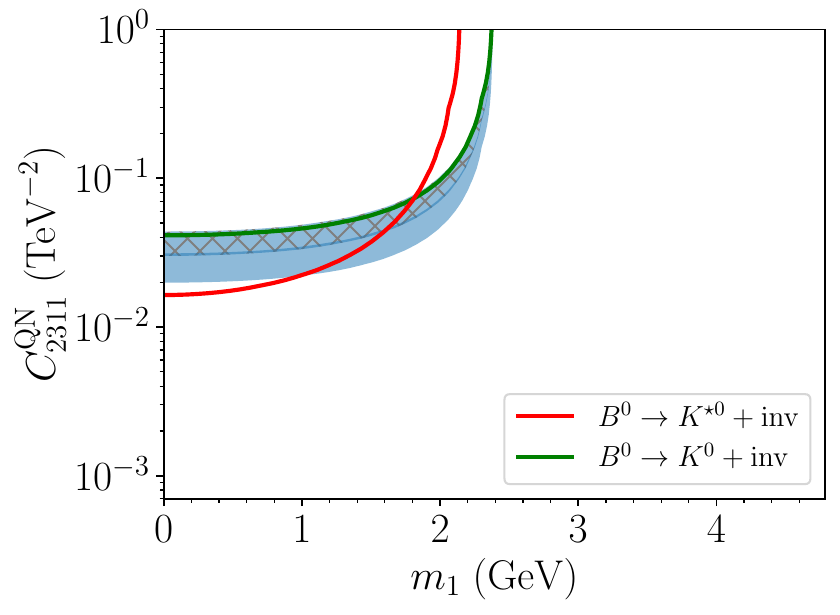}%
    \includegraphics[width=0.5\textwidth]{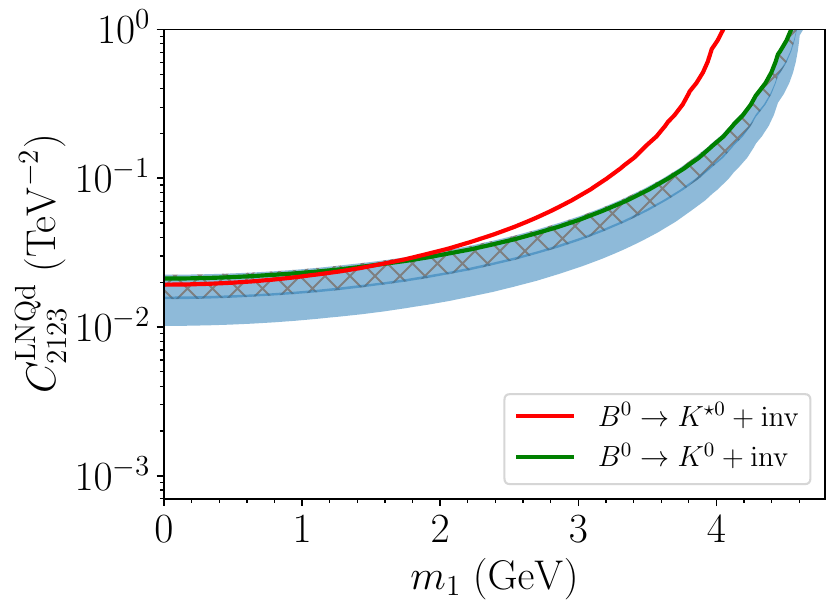}%
 
    \begin{minipage}[c]{0.5\textwidth}
    \includegraphics[width=\textwidth]{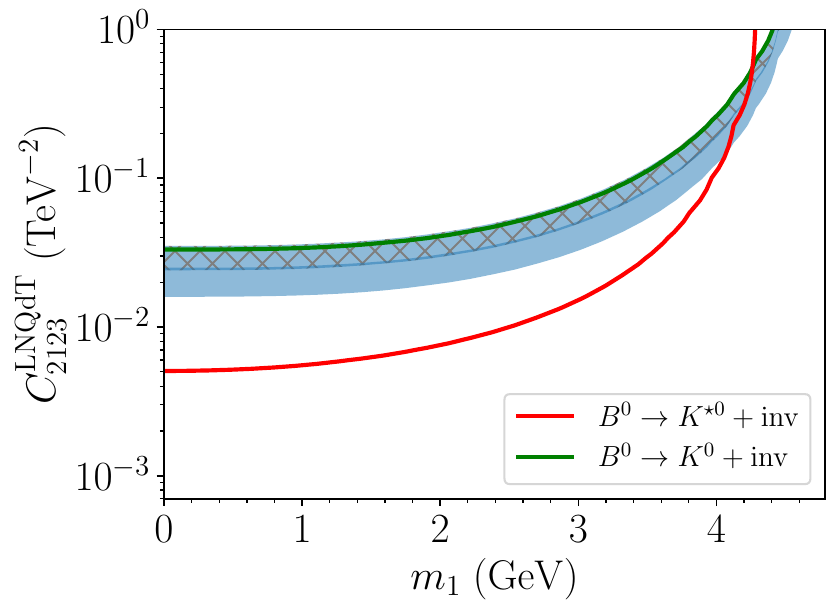}%

    \end{minipage}\hfill
	\begin{minipage}[c]{0.45\textwidth}
    
    \caption{The red (green) contour line stands for the present bound on BR($B^0\to K^{\star 0}(K^0)+\mathrm{inv}$), and the respective regions above these lines are therefore ruled out. 
    The light-blue band symbolises the simple weighted average for BR($B^+\to K^++\mathrm{inv}$), and the hatched light-blue region is compatible with the 2023 Belle II measurement.
    }
    \label{fig:CvsM}
    \end{minipage}
\end{figure*}

Although the charged current processes are well constrained, they do not provide competitive constraints. In our analysis we find that they are always subdominant compared to other rare $B$ meson decays $B\to K^{(\star)}+\mathrm{inv}$ and are thus not included in the discussion of Sec.~\ref{sec:results}. This can be understood from the different sizes of the SM contributions to the different processes: While charged current processes are induced at tree level in the SM via $W$ boson exchange, rare decays are only induced at loop level and are further suppressed by the Glashow-Iliopoulos-Maiani (GIM) mechanism. The new physics contributions from the operators in Eq.~\eqref{eq:Lagrangian} are however of the same size for both charged-current and neutral-current processes.
Finally, colliders may be used to search for new charged-current semi-leptonic operators, see e.g.~\cite{Greljo:2018tzh,Han:2022uho}, which results in constraints on the new physics scale in the $O(1-10)$ TeV range. Long-lived sterile neutrinos from meson decays may also be searched for at the proposed far-forward detectors at the (high-luminosity) LHC, SHiP and Belle II, see e.g.~\cite{DeVries:2020jbs,Zhou:2021ylt,Beltran:2022ast}. There are currently no constraints from leptonic $B_c$ decays. 

\section{Results}
\label{sec:results}

In the following we focus on the rare $B$ meson decay $B\to K^{(\star)} +\mathrm{inv}$ and present first an analysis with one operator being switched on at a time, which is followed by a discussion of correlations between pairs of operators.

In Fig.~\ref{fig:CvsM} we illustrate how a single operator may explain the observed decay $B^+\to K^++\mathrm{inv}$ as a function of the sterile neutrino mass $m_1$. The light-blue band indicates the $1\sigma$-allowed region of parameter space which explains the simple weighted average for BR($B^+\to K^++\mathrm{inv}$). The hatched light-blue region indicates the parameter space explaining the Belle II measurement. Wilson coefficients above the solid red and green contour lines are excluded at 90\% CL by the non-observation of $B^0\to K^{\star0} +\mathrm{inv}$ and $B^0\to K^0 +\mathrm{inv}$, respectively. 
The results do not depend on the neutrino flavour. In particular, the results are unchanged, when changing from first generation lepton doublets to second or third generation lepton doublets. We do not show plots for the scalar and tensor operators with a right-handed strange quark. They yield identical results as the operators with a left-handed strange quark, because the branching ratios are invariant under the exchange $s\leftrightarrow b$ for the LEFT scalar and tensor operators. 

\begin{figure*}[ptb!]
    \centering
     \includegraphics[width=0.5\textwidth]{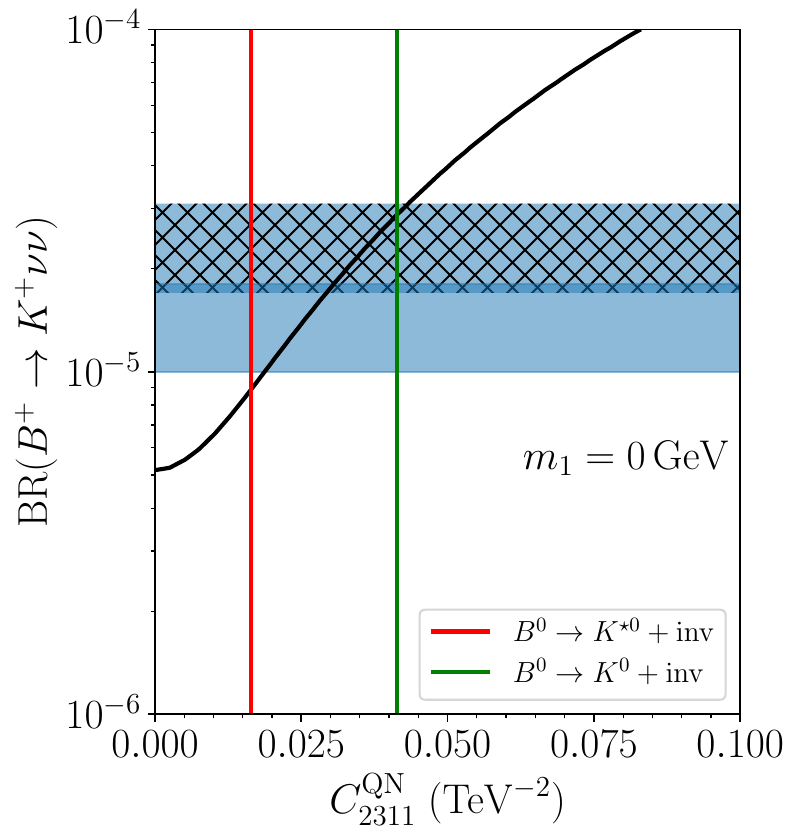}%
     \includegraphics[width=0.5\textwidth]{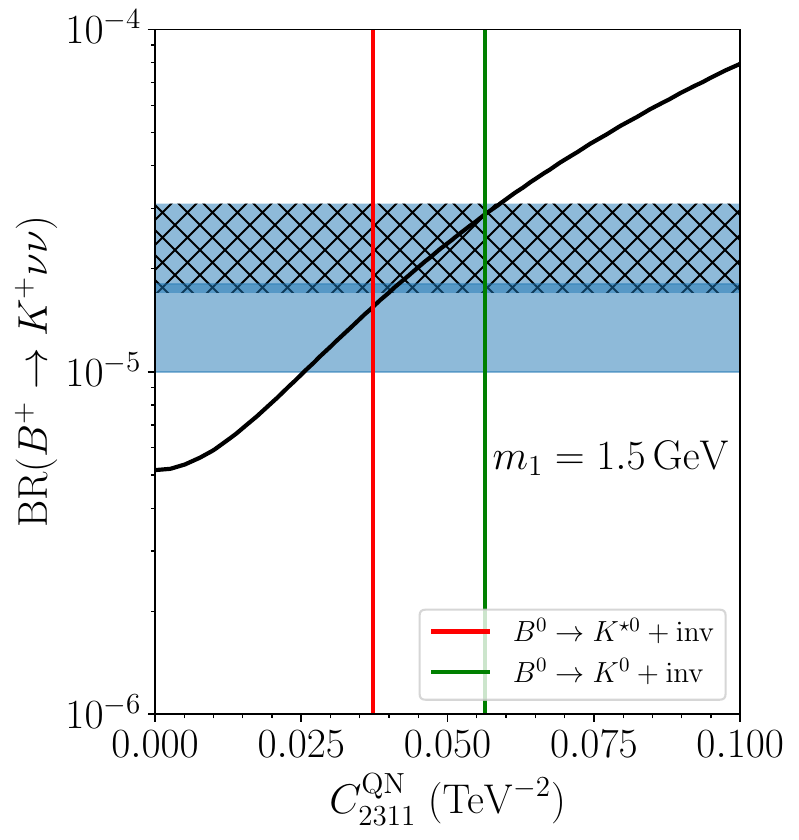}%
     
     \includegraphics[width=0.5\textwidth]{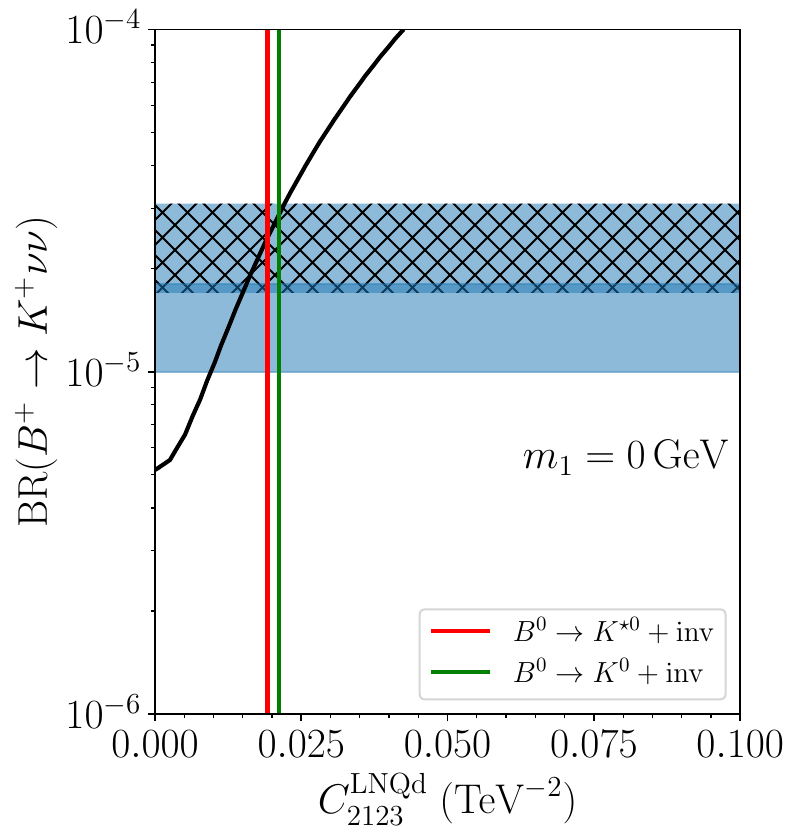}%
     \includegraphics[width=0.5\textwidth]{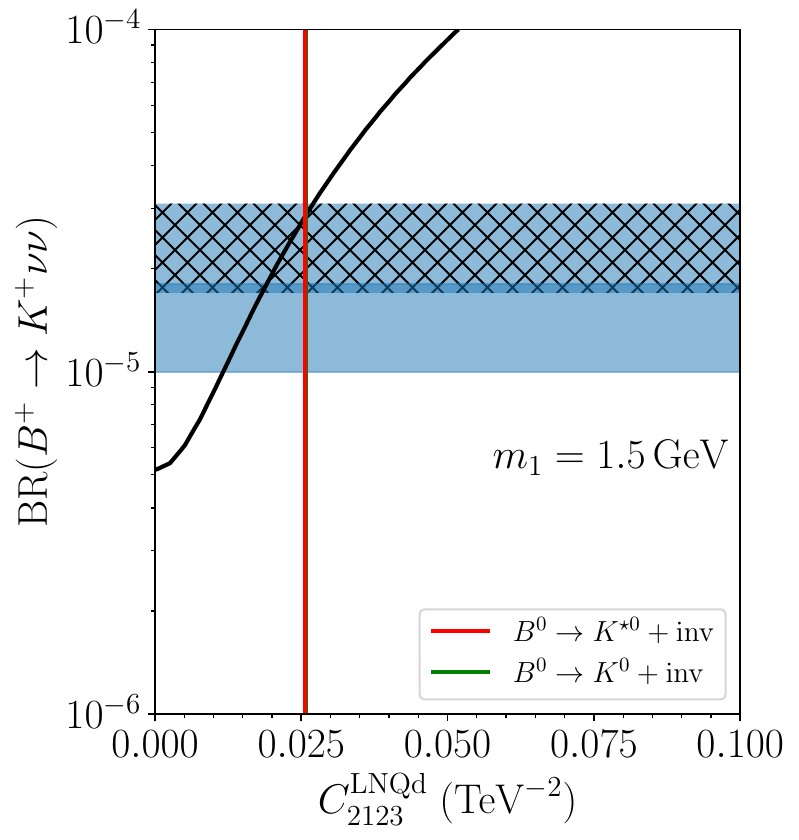}
    \caption{The red (green) line stands for the present bound on BR($B^0\to K^{\star 0}(K^0)+\mathrm{inv}$), and the respective regions to the right of these lines are therefore ruled out. The light-blue band symbolises the simple weighted average for BR($B^+\to K^++\mathrm{inv}$), and the hatched light-blue region is compatible with the 2023 Belle II measurement. The plots in the top row also apply to the operator $\mathcal{O}^{\rm dN}$ and the plots in the bottom row also apply to $\mathcal{O}^{\rm LNQd}_{2132}$.}
    \label{fig:BR_B2Kpnunu}
\end{figure*}
Fig.~\ref{fig:CvsM} (bottom) clearly shows that tensor operators are not able to explain the observed excess in $B^+\to K^++\mathrm{inv}$. The constraint from $B\to K^\star +\mathrm{inv}$ is weakened for large sterile neutrino masses $m_1\sim 4$ GeV because of the larger mass of the $K^\star$ meson and thus reduced phase space. As large sterile neutrino masses modify the $q^2$ distribution and the experimental efficiency is generally lower for large $q^2$, a more careful analysis using the full $q^2$ distribution is required in this case. 
The result is independent of the lepton flavour and the same result applies to the tensor operator with right-handed strange quarks. 

While vector operators $\mathcal{O}^{\rm QN}$ with massless sterile neutrinos are clearly excluded as explanation, Fig.~\ref{fig:CvsM} (top left) suggests the ability to explain the observed excess for sterile neutrino masses of $m_1\gtrsim 1.5$ GeV. This feature is again due to the smaller phase space for decays to $K^\star$ mesons and requires a more detailed analysis using the $q^2$ distribution to arrive at a conclusive result. The same conclusion holds for the vector operator $\mathcal{O}^{\rm dN}$ with right-handed down-type quarks.  

As shown in Fig.~\ref{fig:CvsM} (top right), the most promising explanation of the observed excess is provided by scalar operators $\mathcal{O}^{\rm LNQd}$ whose contribution to $B\to K^\star \nu\nu$ is suppressed. This opens up the possibility to explain the observed excess for the whole allowed sterile neutrino mass range. Note, although scalar operators with massless neutrinos do not have a sharp cutoff in the $q^2$ distribution, they dominantly contribute to the large $q^2$ region as illustrated in Fig.~\ref{fig:diff}. As it has been pointed out in~\cite{Bause:2021cna,Bause:2023mfe}, the Belle II experiment is able to constrain them by searching for invisible $B_s$ decays which are currently unconstrained. 

\begin{figure*}[ptb!]
    \centering
     \includegraphics[width=0.5\textwidth]{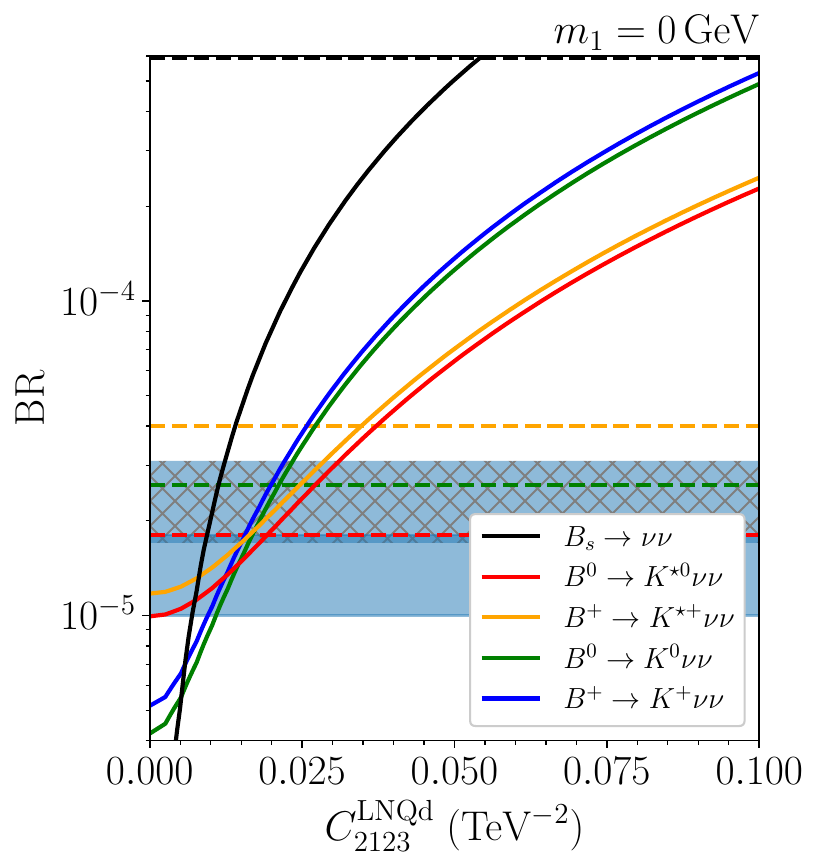}%
     \includegraphics[width=0.5\textwidth]{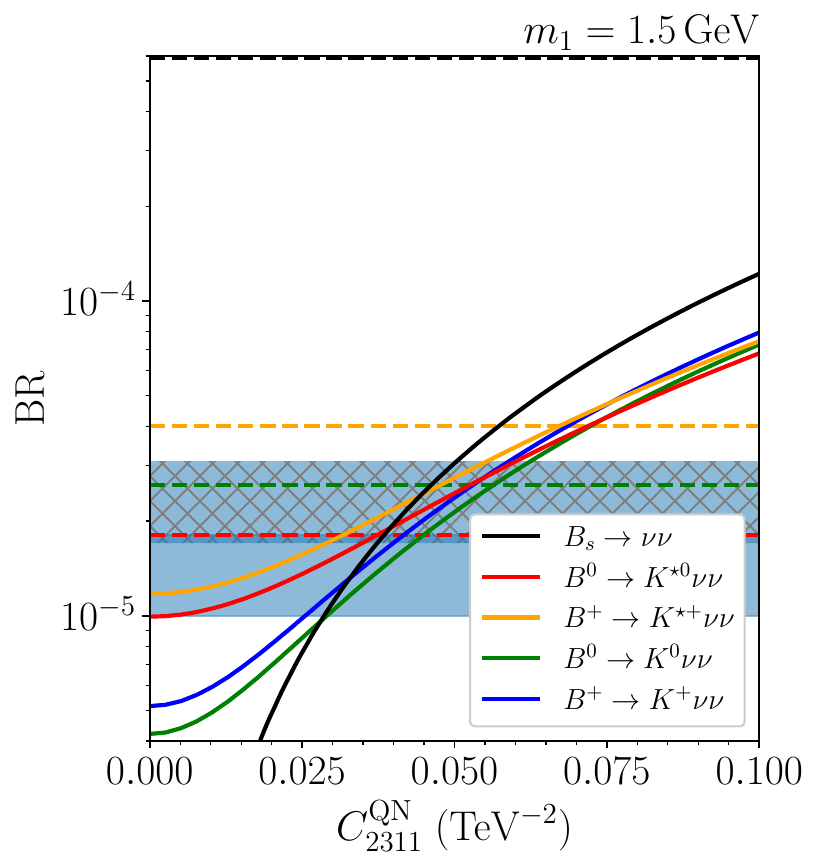}
    \caption{Branching ratios of several decay modes as a function of the scalar (left) and vector (right) Wilson coefficients for fixed sterile neutrino mass. The dashed horizontal contours indicate the current upper bounds for $B^0\to K^{\star0} +\mathrm{inv}$ (red),  $B^+\to K^{\star+}+\mathrm{inv}$ (orange), $B^0\to K^0+\mathrm{inv}$ (green), and $B_s\to \mathrm{inv}$ (black). The light-blue band symbolises the simple weighted average for BR($B^+\to K^++\mathrm{inv}$) and the hatched light-blue region is compatible with the 2023 Belle II measurement.}
    \label{fig:BR-WC}
\end{figure*}

In the following we focus on scalar and vector operators which are most promising to explain the observed excess in $B^+\to K^++\mathrm{inv}$.
Fig.~\ref{fig:BR_B2Kpnunu} shows the dependence of the (total) branching ratio BR($B^+\to K^+\nu\nu$) as a function of the Wilson coefficient for two benchmark masses of $m_1=0\,(1.5)$ GeV on the left (right). The vertical red (green) line indicates the most stringent constraint from $B\to K^\star +\mathrm{inv}$ ($B\to K + \mathrm{inv}$), which excludes the region to the right of the line. The horizontal light-blue bands again indicate the region preferred by $B^+\to K^++\mathrm{inv}$. 
Fig.~\ref{fig:BR_B2Kpnunu} (top row) shows the branching ratio as a function of the Wilson coefficient for the vector operator $\mathcal{O}^{\rm QN}_{2311}$ with left-handed quark doublets.  The top-left figure clearly illustrates the conflict between the measurement of $B^+\to K^++\mathrm{inv}$ and the non-observation of $B^0\to K^{\star0}+\mathrm{inv}$ for massless sterile neutrinos. For $m_1=1.5$ GeV, $B^0\to K^{\star0}+\mathrm{inv}$ constrains $C_{2311}^{\rm QN}\leq 0.037\,\mathrm{TeV}^{-2}$, which limits the excess in $B^+\to K^++\mathrm{inv}$ to the region allowed by the simple weighted average, but precludes an explanation of the larger branching ratio measured by Belle II. 
The same result applies for the vector operator $\mathcal{O}^{\rm dN}$ with right-handed down-type quarks.
Similarly, Fig.~\ref{fig:BR_B2Kpnunu} (bottom row) illustrates the dependence of BR($B^+\to K^+\nu\nu$) on the scalar Wilson coefficient $C^{\rm LNQd}_{2123}$. The region to the left of the vertical green and red lines is consistent with the neutral decay modes $B^0\to K^{\star0}\nu\nu$ and thus the scalar operators are able to explain the observed excess. 

In order to compare how future experimental results may impact the conclusions, we show in 
Fig.~\ref{fig:BR-WC} the branching ratios for invisible $B_s$ decay (black), $B^0\to K^{\star0}\nu\nu$ (red), $B^+\to K^{\star +}\nu\nu$ (orange), $B^0\to K^0\nu\nu$ (green), and $B^+\to K^+\nu\nu$ (blue) as a function of the scalar $C^{\rm LNQd}_{2123}$ and vector $C^{\rm QN}_{2311}$ Wilson coefficient as solid lines and the current upper limits as dashed lines. The region to the right of the intersection of the dashed and solid lines of the same colour is already excluded. The branching ratios of the charged and neutral $B$ meson decay channels are approximately equal due to isospin symmetry. Belle II provided a detailed study of future sensitivities in~\cite{Belle-II:2022cgf}: In particular with an integrated luminosity of $1\,\mathrm{ab}^{-1}$, Belle II is sensitive to the signal strength of $2.06$ for $B^0\to K^0 +\mathrm{inv}$ and thus is able to probe the entire region favoured by $B^+\to K^++\mathrm{inv}$. An improved analysis may even reach a signal strength of $1.37$. Similarly, the expected sensitivities for the decays $B\to K^\star +\mathrm{inv}$ are $1.08 (0.72)$ and $2.04 (1.45)$ for the neutral and charged $B$ meson decays, where we indicated the expected improved sensitivities in parentheses. Hence, also the searches for $B^0\to K^{\star0}+\mathrm{inv}$ are able to probe most of the preferred parameter space for scalar operators, while $B^+\to K^{\star+}+\mathrm{inv}$ is less sensitive. For the vector operator explanation, the decays $B\to K^\star \nu\nu$ are not expected to be sensitive to the whole preferred parameter space.

There is currently only a weak upper limit on the branching ratio of invisible $B_s$ decays based on LEP data~\cite{Alonso-Alvarez:2023mgc}.
Belle II's ability to probe BR($B_s\to \mathrm{inv})\gtrsim 1.1\times 10^{-5}$ with an integrated luminosity of $5\, \mathrm{ab}^{-1}$~\cite{Belle-II:2018jsg} will provide a conclusive test of the described explanations of $B^+\to K^+\nu\nu$ for both scalar and vector operators. The scalar operator contribution is not helicity suppressed, as outlined in Sec.~\ref{sec:pheno}, and thus the invisible $B_s$ decay is sensitive to the full sterile neutrino mass range. The vector operator contribution is helicity suppressed, but the explanation of the excess in $B^+\to K^++\mathrm{inv}$ requires large sterile neutrino masses $m_1\gtrsim 1.5$ GeV and thus invisible $B_s$ decays are also able to probe the vector operator scenario.

\begin{figure*}[htb!]
    \centering
    
    \includegraphics[width=0.5\textwidth]{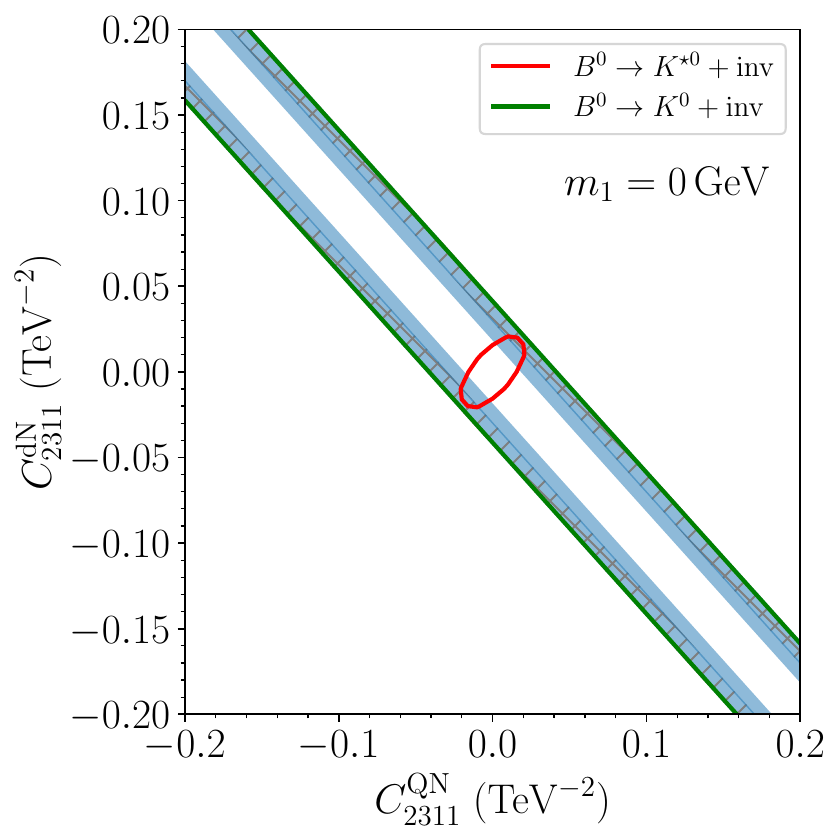}%
    \includegraphics[width=0.5\textwidth]{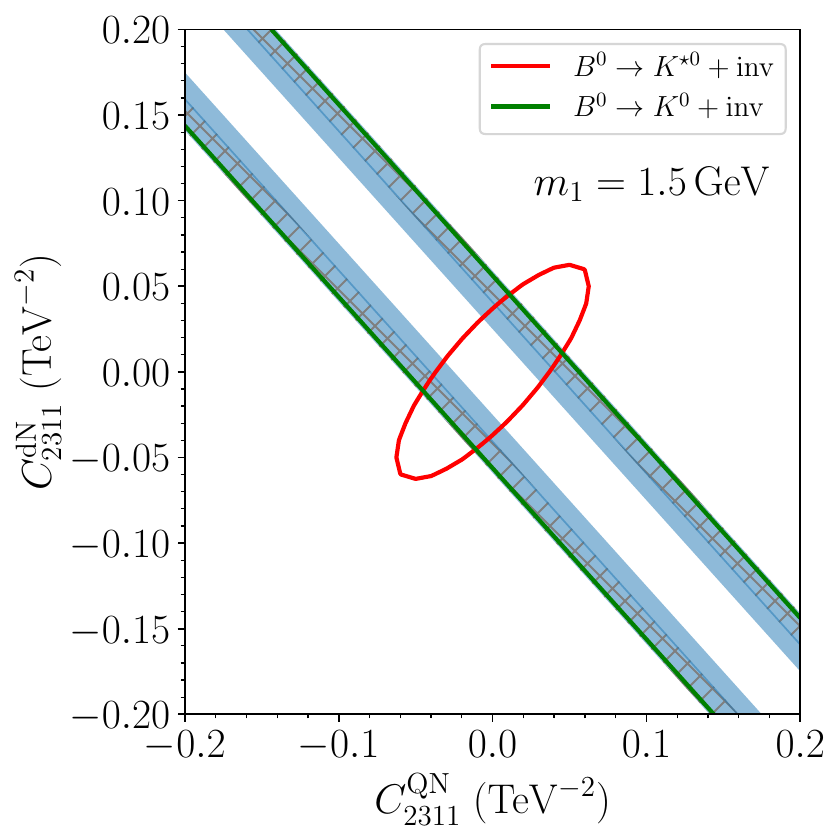}

    \includegraphics[width=0.5\textwidth]{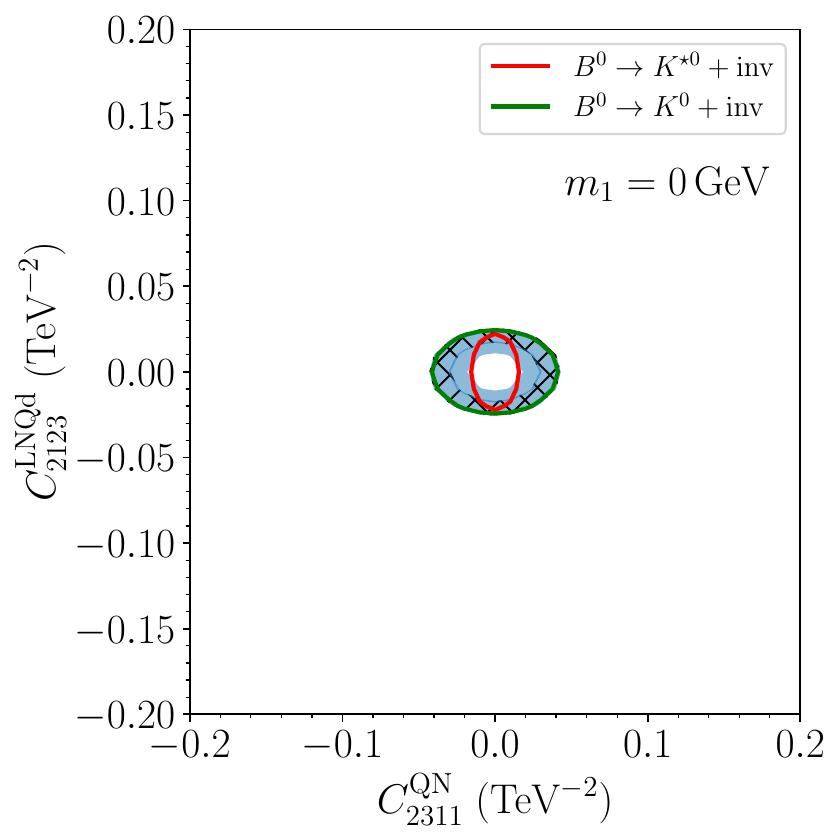}%
    \includegraphics[width=0.5\textwidth]{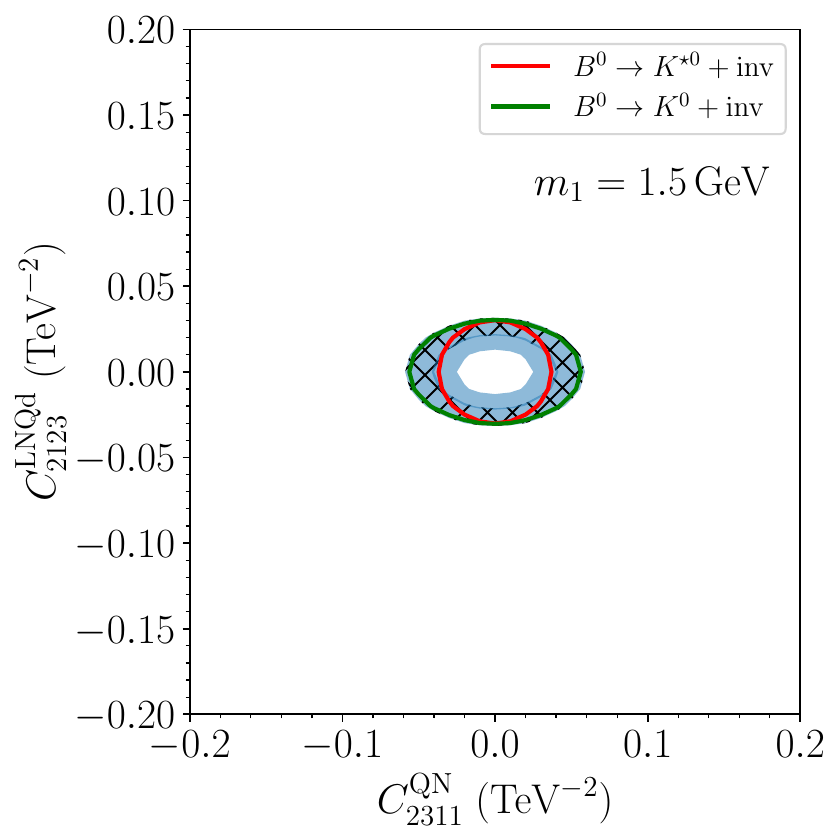}
    
    \caption{The red (green) contour line stands for the present bound on BR($B^0\to K^{\star 0}(K^0)+\mathrm{inv}$), and the respective regions outside of these lines are therefore ruled out. The light-blue band symbolises the simple weighted average for BR($B^+\to K^++\mathrm{inv}$), and the hatched light-blue region is compatible with the 2023 Belle II measurement.
    }
    \label{fig:correlations}
\end{figure*}

We finally turn our attention to correlations between different operators.
In Fig.~\ref{fig:correlations}, we present the 90\% CL exclusion contours for $B^0\to K^{\star 0} +\mathrm{inv}$ ($B^0\to K^0 +\mathrm{inv}$) in red (green) using the current upper limits~\cite{Belle:2013tnz,Belle:2017oht}. Larger Wilson coefficients are excluded. The constraint from $B^+\to K^{\star+}+\mathrm{inv}$ is weaker than $B^0\to K^{\star 0}+\mathrm{inv}$ and thus not shown for clarity. Similar to above, the blue-shaded region indicates the simple weighted average for BR($B^+\to K^++\mathrm{inv}$) and the hatched region the 2023 Belle II measurement of $B^+\to K^++\mathrm{inv}$. 

Fig.~\ref{fig:correlations} (top row) illustrates the correlation between the vector operators with right-handed down-type quarks, $\mathcal{O}^{\rm dN}_{2311}$, and left-handed quark doublets, $\mathcal{O}^{\rm QN}_{2311}$. Their contribution to the branching ratio 
of $B\to K\nu\nu$ is proportional to $\left|C^{\rm QN}_{2311} + C^{\rm dN}_{2311}\right|^2$, while for the branching ratios of $B\to K^\star\nu\nu$ there are two independent contributions proportional to the square of the sum and difference of the two Wilson coefficients. This explains the shape of the contour lines: While the non-observation of $B\to K^\star +\mathrm{inv}$ results in an elliptical exclusion region, the decay $B\to K\nu\nu$ is insensitive to $C^{\rm QN}_{2311} = -C^{\rm dN}_{2311}$ and thus the search for $B\to K+\mathrm{inv}$ is unable to probe this direction in parameter space. Increasing the sterile neutrino mass to $m_1=1.5$ GeV, see Fig.~\ref{fig:correlations} (top right), reduces the contribution to $B\to K^\star \nu\nu$ and thus weakens the $B\to K^\star+\mathrm{inv}$ constraint. Consequently, the allowed region inside the red ellipse widens.

The plots in Fig.~\ref{fig:correlations} (bottom row) show the correlation between the scalar operator $\mathcal{O}^{\rm LNQd}_{2123}$ and the vector operator $\mathcal{O}^{\rm QN}_{2311}$ with left-handed quark doublets. The elliptical exclusion regions are due to the absence of complete cancellations between the two operators. In fact, there is no interference between the two operators in the massless limit. The bottom-left plot clearly shows the preference for the pure scalar operator scenario for massless sterile neutrinos. For massive sterile neutrinos, the constraint from $B\to K^\star+\mathrm{inv}$ weakens. The bottom-right plot shows the case with $m_1=1.5$ GeV, where the dominant constraint originates from $B^0\to K^{\star0}+\mathrm{inv}$ and the constraint from $B\to K^0+\mathrm{inv}$ is subdominant. The same conclusions holds, if the quark flavours of the scalar operator are exchanged, $s\leftrightarrow b$, or if the vector operator $\mathcal{O}^{\rm dN}$ with right-handed down-type quarks is considered.

\section{UV completions}
\label{sec:uv_completions}
We briefly outline possible UV completions for the scalar and vector operators discussed in the previous section, see \cite{Beltran:2023ymm} for a detailed discussion of tree-level UV completions of dimension-6 operators in $\nu$SMEFT. The most straightforward UV completion of the scalar operator $\mathcal{O}^{\rm LNQd}$ is provided by a second electroweak doublet scalar $\eta\sim (\mathbf{1},\mathbf{2},-\tfrac12)$ with its Yukawa interactions $\bar L N \eta$ and $\bar Q i\sigma_2\eta^* d$. Integrating out the electroweak doublet scalar $\eta$ results in the scalar operator $\mathcal{O}^{\rm LNQd}$, but it also generates the 4-fermion operators $(\bar L \gamma_\mu L) (\bar N \gamma^\mu N)$ and $(\bar Q \gamma_\mu Q) (\bar d \gamma_\mu d)$. The latter contributes to $B_s-\overline{B_s}$ mixing and the former to charged lepton decays $\ell_i\to \ell_j +\mathrm{inv}$ which both have to be considered in a UV-complete model. The required size of the scalar Wilson coefficient $C^{\rm LNQd}\sim 0.02\, \mathrm{TeV}^{-2}$ requires the electroweak doublet scalar to have a mass of at most $\mathcal{O}(7)$ TeV for Yukawa couplings of order unity.

The vector operators are straightforwardly generated using leptoquarks, see e.g.~\cite{Dorsner:2016wpm} for a review. The $S_1\sim(\mathbf{\bar 3},\mathbf{1},\tfrac13)$ leptoquark leads to the operator $\mathcal{O}^{\rm dN}$ via its Yukawa interaction $\overline{N^c} S_1 d$ and thus does not contribute to $B_s-\overline{B_s}$ mixing at tree level. 
Its other leptoquark couplings generate the SMEFT operators $\mathcal{O}^{\rm eu}$, and $\mathcal{O}^{\rm LeQu(1,3)}$, and $\mathcal{O}^{\rm LQ(1,3)}$, see e.g.~\cite{deBlas:2017xtg}.  Similarly, the leptoquark $\tilde R_2\sim (\mathbf{3},\mathbf{2},\tfrac16)$ may generate the operator $\mathcal{O}^{\rm QN}$ via its Yukawa interaction $\overline{Q} \tilde R_2 N$, which does not contribute to $B_s-\overline{B_s}$ mixing at tree level. 
Its other Yukawa interaction $\bar d \tilde R_2 L$ contributes to the SMEFT operator $\mathcal{O}^{\rm Ld}$~\cite{deBlas:2017xtg}. The required size of the vector Wilson coefficient $C^{\rm QN,dN}\sim 0.03\, \mathrm{TeV}^{-2}$ requires the leptoquark to have a mass of at most $\mathcal{O}(6)$ TeV for Yukawa couplings of order unity.
The operators $\mathcal{O}^{\rm LQ(1,3)}$ and $\mathcal{O}^{\rm Ld}$ may also contribute to $B\to K^{(\star)}\nu\nu$, but they are strongly constrained by searches for $B\to K^{(\star)} \ell\bar\ell$ for light charged leptons $\ell$, see e.g.~\cite{Bause:2021cna,Bause:2023mfe}.

\section{Conclusions}
\label{sec:conclusions}

We studied how the recently observed excess in $B^+\to K^++\mathrm{inv}$ may be explained within $\nu$SMEFT. Within $\nu$SMEFT there are four 4-fermion operators with sterile neutrinos which may contribute to $B^+\to K^++\mathrm{inv}$: two vector, one scalar and one tensor operator. We would like to stress that massive sterile neutrinos and scalar and tensor operators modify the $q^2$ distribution with respect to the SM prediction based on massless neutrinos contributing via a vector operator. The present analysis is entirely based on the branching ratios and we only comment on the $q^2$ distribution, because it is currently not available. The tensor operator is strongly constrained by  $B^0\to K^{\star 0} +\mathrm{inv}$ and thus is not able to significantly contribute to $B^+\to K^+ +\mathrm{inv}$. 

We find that the excess in $B^+\to K^+ +\mathrm{inv}$ is most straightforwardly explained by scalar operators with sterile neutrinos irrespective of their mass. This scenario can be conclusively probed using invisible $B_s$ decays at Belle II and the rare semi-leptonic decay $B^0\to K^{(\star) 0}+\mathrm{inv}$ in the near future using $1\,\mathrm{ab}^{-1}$ of data. 
Vector operators may be able to explain the excess of $B^+\to K^+ + \mathrm{inv}$ for intermediate sterile neutrino masses of $m_1\gtrsim 1.5$ GeV assuming no significant changes to the $q^2$ distribution. While  $B\to K^\star +\mathrm{inv}$ strongly constrains this possibility for light sterile neutrinos, the decay $B\to K^\star +\mathrm{inv}$ is less sensitive for sterile neutrino masses $m_1\sim 1.5$ GeV due to phase space suppression, but still provides a constraint. Most of the parameter space of this solution can similarly be probed with the invisible $B_s$ decay and $B^0\to K^{(\star)0}+\mathrm{inv}$. 

We finally outline tree-level UV completions for the scalar and vector operators. The scalar operator may be generated using an electroweak doublet scalar, which however also contributes to $B_s-\overline{B_s}$ mixing at tree level. The vector operators can be straightforwardly generated using leptoquarks: the $S_1$ leptoquark may result in $\mathcal{O}^{\rm dN}$ and the $\tilde R_2$ leptoquark in $\mathcal{O}^{\rm QN}$. Generally, all UV completions also generate other $\nu$SMEFT operators, but for the two leptoquarks it is in principle possible to avoid them. It however requires to set all other Yukawa couplings to zero, which may be considered unnatural. This motivates to consider the full set of relevant $\nu$SMEFT operators in a future study and to investigate explanations of the excess in UV complete models together with other observables.

\section*{Acknowledgements}
MS would like to thank Yi Cai for useful discussions. 
AG and RM would like to thank Michael A. Schmidt for the kind hospitality during their visit to UNSW funded by the Gordon Godfrey bequest.
All figures have been produced using \texttt{matplotlib}~\cite{Hunter:2007}. 
MS and TF acknowledge support by the Australian Research Council Discovery Project DP200101470. 

\appendix

\mathversion{bold}
\section{\texorpdfstring{$S,P,V,A,\mathcal{T}$}{S,P,V,A,T} basis}
\mathversion{normal}
\label{sec:SPVAT}
For the calculation it is convenient to employ a basis with well-defined parity properties which has been used in~\cite{Gratrex:2015hna}. In this appendix we provide the matching of the chiral LEFT basis used in the main part of the text to this basis both for charged current operators with Dirac fermions and neutral current neutrino-quark operators with Majorana fermions. 

\subsection{Charged current operators with Dirac fermions}
We define the operator basis as
\begin{align}\label{eq:Lagrangian_Gratrex}
    \mathcal{L} &= c_H \sum_i\sum_{\alpha,\beta} (C_{i,\alpha\beta} O_{i,\alpha\beta} 
    +C_{i,\alpha\beta}^\prime O_{i,\alpha\beta}^\prime)
    \;,
\end{align}
where $i$ runs over $S,P,V,A,\mathcal{T}$ and $c_H$ determines the normalisation of the operators. Similar to \cite{Felkl:2021uxi} we use $c_H=1$, which differs from \cite{Gratrex:2015hna} where the normalisation $c_H=\frac{4G_F}{\sqrt{2}} \frac{\alpha}{4\pi} V_{ts}^* V_{tb}$ has been used. The operators are 
\begin{equation}
\label{eq:SPVAT-Dirac}
    \begin{aligned}
    O_{S(P)\alpha\beta} & = (\overline{c_L} b)  (\overline{e_\alpha}(\gamma_5)\nu_\beta)\;,
                        & 
    O_{V(A)\alpha\beta} & = (\overline{c_L} \gamma^\mu b) (\overline{e_\alpha}\gamma_\mu (\gamma_5)\nu_\beta)\;,
                        &
    O_{\mathcal{T}\alpha\beta} & = (\overline{c_L} \sigma^{\mu\nu} b) (\overline{e_\alpha} \sigma_{\mu\nu}\nu_\beta)
    \;.
    \end{aligned}
\end{equation}
The primed operators are obtained by replacing
$c_L\to c_R$, i.e. $\mathcal{O}^\prime=\mathcal{O}|_{
c_L\to c_R}$ where $q_{L,R}\equiv P_{L,R}q$. The Wilson coefficients in the $S,P,V,A,\mathcal{T}$ basis are given in terms of the chiral basis by
\begin{equation}
\begin{aligned}
    C_{V\alpha\beta} & = \frac{1}{2}\left(C_{\nu edu,\beta\alpha bc}^{\text{VRL}*} + C_{\nu edu,\beta\alpha bc}^{\text{VLL}*}\right) \;,
                     &
    C_{V\alpha\beta}^\prime & =  \frac{1}{2}\left(C_{\nu edu,\beta\alpha bc}^{\text{VRR}*} + C_{\nu edu,\beta\alpha bc}^{\text{VLR}*}\right) \;,
    \\
    C_{A\alpha\beta} & = \frac{1}{2}\left(C_{\nu edu,\beta\alpha bc}^{\text{VRL}*} - C_{\nu edu,\beta\alpha bc}^{\text{VLL}*}\right) \;,
                            &
    C_{A\alpha\beta}^\prime & = \frac{1}{2}\left(C_{\nu edu,\beta\alpha bc}^{\text{VRR}*} - C_{\nu edu,\beta\alpha bc}^{\text{VLR}*}\right) \;,
    \\
    C_{S\alpha\beta} & =  \frac{1}{2}\left(C_{\nu edu,\beta\alpha bc}^{\text{SLL}*} + C_{\nu edu,\beta\alpha bc}^{\text{SRL}*}\right) \;,
                     &
    C_{S\alpha\beta}^\prime & =  \frac{1}{2}\left(C_{\nu edu,\beta\alpha bc}^{\text{SLR}*} + C_{\nu edu,\beta\alpha bc}^{\text{SRR}*}\right) \;,
    \\
    C_{P\alpha\beta} & = \frac{1}{2}\left(C_{\nu edu,\beta\alpha bc}^{\text{SLL}*} - C_{\nu edu,\beta\alpha bc}^{\text{SRL}*}\right) \;,
                            &
    C_{P\alpha\beta}^\prime & = \frac{1}{2}\left(C_{\nu edu,\beta\alpha bc}^{\text{SLR}*} - C_{\nu edu,\beta\alpha bc}^{\text{SRR}*}\right) \;,
    \\
    C_{T\alpha\beta} & = C_{\nu edu,\beta\alpha bc}^{\text{TLL}*}\;,
                     &
    C_{T\alpha\beta}^\prime & = C_{\nu edu,\beta\alpha bc}^{\text{TRR}*}\;.
\end{aligned}
\end{equation}

\subsection{Neutral current neutrino-quark operators with Majorana neutrinos}
Following \cite{Felkl:2021uxi} we define the leptonic helicity amplitudes for Majorana spinors with an additional factor of $1/2$ in the effective Lagrangian, such that the leptonic helicity amplitudes have the same form as for Dirac fermions.
 The effective Lagrangian is thus given by 
\begin{align}
    \mathcal{L} &= \frac12 c_H \sum_i\sum_{\alpha,\beta} (C_{i,\alpha\beta} O_{i,\alpha\beta} 
    +C_{i,\alpha\beta}^\prime O_{i,\alpha\beta}^\prime)
    \;.
\end{align}
The operators with Majorana neutrinos are obtained from Eq.~\eqref{eq:SPVAT-Dirac} using $e_\alpha\to \nu_\alpha$ and $c\to s$. The operators have well-defined symmetry properties: the (pseudo-)scalar and axial-vector operators are symmetric in the neutrino flavour indices and the vector and tensor operators are antisymmetric.
We find for the Wilson coefficients using the $S,P,V,A,\mathcal{T}$ basis 
\begin{equation}
\begin{aligned}
    C_{V\alpha\beta} & = C_{\nu d,[\alpha\beta] sb}^{\text{VLL}} \;,
                     &
    C_{V\alpha\beta}^\prime & = C_{\nu d,[\alpha\beta] sb}^{\text{VLR}} \;,
    \\
    C_{A\alpha\beta} & = - C_{\nu d,(\alpha\beta) sb}^{\text{VLL}} \;,
                     &
    C_{A\alpha\beta}^\prime & = - C_{\nu d,(\alpha\beta) sb}^{\text{VLR}}  \;,
 \\
 C_{S\alpha\beta} & =  C_{\nu d,(\alpha\beta) sb}^{\text{SLR}} +  C_{\nu d,(\beta\alpha) bs}^{\text{SLL}*} \;,
                  &
 C_{S\alpha\beta}^\prime & =  C_{\nu d,(\alpha\beta) sb}^{\text{SLL}} + C_{\nu d,(\beta\alpha) bs}^{\text{SLR}*}\;,
 \\
 C_{P\alpha\beta} & = - C_{\nu d,(\alpha\beta) sb}^{\text{SLR}} +  C_{\nu d,(\beta\alpha) bs}^{\text{SLL}*}\;,
                  &
 C_{P\alpha\beta}^\prime & = - C_{\nu d,(\alpha\beta) sb}^{\text{SLL}} + C_{\nu d,(\beta\alpha) bs}^{\text{SLR}*}\;,
 \\
 C_{\mathcal{T}\alpha\beta} & =  2C_{\nu d,[\beta\alpha] bs}^{\text{TLL}*} \;,
                            &
 C_{\mathcal{T}\alpha\beta}^\prime & = 2 C_{\nu d, [\alpha\beta] sb}^{\text{TLL}} \;,
\end{aligned}
\end{equation}
where $\alpha,\beta$ denote the neutrino flavours.
Parentheses (\dots) indicate symmetrisation and square brackets [\dots] indicate anti-symmetrisation of the neutrino flavour indices as in
\begin{align}
    M_{(ab)} &\equiv \frac12 \left(M_{ab}+M_{ba}\right) \;,
    &
    M_{[ab]} &\equiv \frac12 \left(M_{ab}-M_{ba}\right)  
    \;.
\end{align}

\section{\texorpdfstring{$B_s\to \nu\nu$}{Bs->nunu}}
\label{sec:Bsnunu}
An important observable is given by invisible $B_s$ decays. In this section, we provide the general branching ratio for $B_s\to\nu_\alpha\nu_\beta$. We define the form factors
\begin{equation} 
\begin{aligned}
    \left\langle 0 | \bar s\gamma^\mu \gamma_5 b | \bar B_s(P)\right\rangle & = if_{B_s} P^\mu\;, 
                                                                            &
    \left\langle 0 | \bar s \gamma_5 b | \bar B_s(P)\right\rangle & = -i\frac{h_{B_s}}{m_b+m_s}\;,
\end{aligned}
\end{equation}
where the scalar and vector form factors are related by $h_{B_s} = m_{B_s}^2 f_{B_s}$. Using the $S,P,V,A,\mathcal{T}$ basis, we find for the branching ratio of $B_s\to \nu_\alpha\nu_\beta$ 
\begin{equation}
\begin{aligned}
  \mathrm{BR}(B_s\to \nu_\alpha\nu_\beta)   &= \frac{\lambda^{1/2}(1,x_\alpha^2,x_\beta^2) \tau_{B_s}m_{B_s}^3 f_{B_s}^2}{64\pi (1+\delta_{\alpha\beta})} 
    \Bigg( 
    \left|C_{V\alpha\beta}-C_{V\alpha\beta}^\prime\right|^2 (1-(x_\alpha+x_\beta)^2)(x_\alpha-x_\beta)^2
    \\&
    +\left|C_{A\alpha\beta}-C_{A\alpha\beta}^\prime\right|^2 (1-(x_\alpha-x_\beta)^2)(x_\alpha+x_\beta)^2
    \\&
    +\left|C_{S\alpha\beta}-C_{S\alpha\beta}^\prime\right|^2 \frac{1-(x_\alpha+x_\beta)^2}{(x_b+x_s)^2}
    +\left|C_{P\alpha\beta}-C_{P\alpha\beta}^\prime\right|^2 \frac{1-(x_\alpha-x_\beta)^2}{(x_b+x_s)^2}
    \\&
    +2\,\mathrm{Re}\left[(C_{V\alpha\beta}-C_{V\alpha\beta}^\prime)(C_{S\alpha\beta}-C_{S\alpha\beta}^\prime)^*\right]
    \frac{x_\alpha(1-x_\alpha x_\beta -x_\alpha^2)-x_\beta(1-x_\alpha x_\beta - x_\beta^2)}{x_b+x_s}
    \\&
    +2\,\mathrm{Re}\left[(C_{A\alpha\beta}-C_{A\alpha\beta}^\prime)(C_{P\alpha\beta}-C_{P\alpha\beta}^\prime)^*\right] 
    \frac{x_\alpha(1+x_\alpha x_\beta -x_\alpha^2)+x_\beta(1+x_\alpha x_\beta - x_\beta^2)}{x_b+x_s}
    \Bigg)\;,
\end{aligned}
\end{equation}
with $x_\alpha = \frac{m_\alpha}{m_{B_s}}$
and the K\"all\'en function $\lambda(x,y,z) \equiv x^2+y^2+z^2-2xy-2xz-2yz$. 
There is no interference between the operators with different neutrino exchange symmetry and there is no contribution from tensor operators. The branching ratio is symmetric in the final state neutrino flavours. 
For identical final state neutrinos the branching ratio reduces to 
\begin{equation}
\begin{aligned}
  \mathrm{BR}(B_s\to \nu_\alpha\nu_\alpha)   =& \frac{\tau_{B_s}m_{B_s}^3 f_{B_s}^2}{128\pi}
    \Bigg( 
    4 \left|C_{A\alpha\alpha}-C_{A\alpha\alpha}^\prime\right|^2 x_\alpha^2\sqrt{1-4x_\alpha^2}
    \\&
    +\left|C_{S\alpha\alpha}-C_{S\alpha\alpha}^\prime\right|^2 \frac{(1-4x_\alpha^2)^{3/2}}{x_b^2}
    +\left|C_{P\alpha\alpha}-C_{P\alpha\alpha}^\prime\right|^2 \frac{\sqrt{1-4x_\alpha^2}}{x_b^2}
    \\&
    +4\,\mathrm{Re}\left[(C_{A\alpha\alpha}-C_{A\alpha\alpha}^\prime)(C_{P\alpha\alpha}-C_{P\alpha\alpha}^\prime)^*\right]
    \frac{x_\alpha \sqrt{1-4x_\alpha^2}}{x_b}
    \Bigg) \;,
\end{aligned}
\end{equation}
where we used $x_s\ll x_b$.

\setlength{\bibsep}{.2\baselineskip plus 0.3ex}
\bibliographystyle{utphys} 
\bibliography{refs}
\end{document}